\documentclass[sigconf]{acmart} 

\usepackage{mdwmath}
\usepackage{mdwtab}

\usepackage{url}
\usepackage{times}

\usepackage{versions}

\usepackage[capitalize]{cleveref}

\usepackage[caption=false,font=footnotesize]{subfig}

\usepackage{paralist}           

\usepackage{algorithmic}       
\usepackage{listings}           
\usepackage{moreverb}           

\usepackage{fancyvrb}           

\usepackage{eurosym}
\usepackage{units}              

\usepackage{booktabs}
\usepackage{csvsimple}          
\usepackage[flushleft]{threeparttable} 



\usepackage{rotating}           

\usepackage{array}
\newcolumntype{L}[1]{>{\raggedright} m{#1}}
\newcolumntype{C}[1]{>{\centering}   m{#1}}
\newcolumntype{R}[1]{>{\raggedleft}  m{#1}}

\usepackage{multirow}

\newcommand{\PreserveBackslash}[1]{\let\temp=\\#1\let\\=\temp}

\usepackage[normalem]{ulem}     

\usepackage{fp}                 

\usepackage{fancyhdr}           

\usepackage{mfirstuc}           

\usepackage{xspace}



\usepackage{todonotes}
\usepackage{xcolor}
\definecolor{pink}{rgb}{1,.4,.4}
\definecolor{red}{rgb}{1, .2, .1}

\newcommand{\WFH}{working-from-home\xspace}
\newcommand{\Wfh}{Working-from-home\xspace}
\newcommand{\workFromHome}{work from home\xspace}
\newcommand{\RQ}{``what changes does a \GSDlong organization need to make for the wellness of their employees during a pandemic?''\xspace}

\newcommand{\Ocuco}{Ocuco Ltd.\xspace}
\newcommand{\SAFelong}{Scaled Agile Framework\textsuperscript\textregistered\xspace}
\newcommand{\SAFe}{SAFe\xspace}

\newcommand{\GSDlong}{global software development\xspace}
\newcommand{\numberofresponses} {seven\xspace}

\renewcommand{\theenumii}{\theenumi.\arabic{enumii}.}

\begin{document}


\pagestyle{fancy}
\newcommand{\preprintfoot}[1]{%
\fancyfoot[l]{\vspace{10pt}\footnotesize{\textcolor{gray}{#1}}}
}

\preprintfoot{This an authors' preprint.  Please cite as: Clodagh
  NicCanna, Mohammad Abdur Razzak, John Noll, and Sarah Beecham (2021) ``Globally Distributed Development during COVID-19''
  \emph{8th International Virtual Workshop on Software Engineering Research and Industrial Practice.}}

\title{Globally Distributed Development during COVID-19}

\FPset{\AuthorCount}{0}         

\author{Clodagh NicCanna}
\email{clodagh.niccanna@ocuco.com}
\affiliation{
  \institution{Ocuco Ltd}
  \city{Dublin}
  \country{Ireland}
}
\FPadd{\AuthorCount}{\AuthorCount}{1}
\FPeval{\clodaghPos}{clip(\AuthorCount)}
\newcommand{\clodaghAuthorPos}{\clodaghPos\xspace}

\author{Mohammad Abdur Razzak}
\email{razzak.abdur@ocuco.com}
\affiliation{
  \institution{Ocuco Ltd}
  \city{Dublin}
  \country{Ireland}
}
\FPadd{\AuthorCount}{\AuthorCount}{1}
\FPeval{\razzakPos}{clip(\AuthorCount)}
\newcommand{\razzakAuthorPos}{\razzakPos\xspace}

\author{John Noll}
\email{j.noll@herts.ac.uk}
\affiliation{
  \institution{University of Hertfordshire}
  \city{Hatfield}
  \state{Herts}
  \country{UK}
}
\FPadd{\AuthorCount}{\AuthorCount}{1}
\FPeval{\nollPos}{clip(\AuthorCount)}
\newcommand{\nollAuthorPos}{\nollPos\xspace}

\author{Sarah Beecham}
\email{sarah.beecham@lero.ie}
\affiliation{
  \institution{Lero, the Irish Software Research Centre}
  \streetaddress{University of  Limerick}
  \city{Limerick}
  \country{Ireland}
}
\FPadd{\AuthorCount}{\AuthorCount}{1}
\FPeval{\beechamPos}{clip(\AuthorCount)}
\newcommand{\beechamAuthorPos}{\beechamPos\xspace}

\ccsdesc[500]{Software and its engineering~Agile software development}
\ccsdesc[500]{Software and its engineering~Programming teams}
\ccsdesc[500]{Social and professional topics~Employment issues}

\renewcommand{\shortauthors}{NicCanna, Noll, Razzak,\& Beecham}

%

\acmConference[SER\&IP]{8th International Virtual Workshop on
Software Engineering Research and Industrial Practice}{June, 2021}{}
\begin{abstract}
Due to the global pandemic, in March 2020 we in academia and industry
were abruptly forced into working from home. Yet teaching never
stopped, and neither did developing software, fixing software, and
expanding into new markets.  Demands for flexible ways of working,
responding to new requirements, have never been so high. \emph{How did
  we manage to continue working, when we had to suddenly switch all
  communication to online and virtual forms of contact?} In this short
paper we describe how Ocuco Ltd., a medium-sized organization headquartered in
Ireland, managed our software development teams -- distributed
throughout Ireland, Europe, Asia and America during the COVID-19
pandemic.  We describe how we expanded, kept our customers happy, and
our teams motivated. We made changes, some large, such as providing
emergency financial support; others small, like implementing regular
online social pizza evenings. Technology and process changes were
minor, an advantage of working in globally distributed teams since
2016, when  development activities were coordinated according to the
Scaled Agile Framework (SAFe). The results of implementing the changes
were satisfying; productivity went up, we gained new customers, and
preliminary results from our wellness survey indicate that everyone
feels extremely well-supported by management to achieve their goals.
However, the anonymised survey responses did show some developers'
anxiety levels were slightly raised, and many are working longer
hours. Administering this survey is very beneficial, as now we know,
so we can act.

\end{abstract}
\keywords{Global Software Development, virtual teams, wellness, remote
  working practices, home-working, \WFH, WFH, pandemic, Covid-19, change management, Scaled Agile Framework}

\maketitle
\thispagestyle{fancy}
\section{Introduction}\label{sec:intro}
The Coronavirus pandemic global crisis has brought academics,
scientists, and practitioners together as we all fight a common cause
for our very survival. A key mitigation strategy to stem the viral
transmission, while balancing health and economic factors, is to work
from home where possible. Software development, especially distributed
software development where teams collaborate across multiple sites, would
seem particularly suited to this transition.  However, even teams
familiar with remote forms or work, would traditionally meet other
members of their teams, face to face, at key points in the project, in
which they would plan for the future, learn together, socialize,
and thereby gain trust~\cite{Noll_2010_Global}. Lack of social contact is
shown to create many additional problems, associated especially with well-being~\cite{Ralph_2020_Pandemic}.

As Europe enters a second and arguably more intense wave of COVID-19
transmissions, in this paper we reflect on how software practitioners
\WFH are faring. 
As a medium-sized enterprise, head-quartered in Ireland,
developing software in distributed teams throughout the globe (see
Figure 1), we share some of our COVID-19 strategies. We recognized
early that changes to our social and technical practices must be made;
to ensure the wellness of all our employees, \emph{acting fast is
key}~\cite{Jones_2020_Optimal}. We heed the warning expressed by a
seasoned developer, who, despite having worked remotely for  many years, after only one month of quarantine noted
``I'm feeling a
tinge of burn out for the first time in my life''~\cite{Ralph_2020_Pandemic}.  

While our experience of
\GSDlong does help, in that we have both infrastructure and process for remote working in
place, this
in itself does not guarantee a smooth transition to \WFH
full time.  Home working presents a very different rhythm and structure
to the working week, and there can be additional pressures of friends and family
needing support, dependents becoming ill, and the worry of reduced
income.
In this paper, we
share our experience with the transition to \WFH, and address
the research question, \RQ

This paper is organized as follows: in the next section we present a
brief background into transitioning from the office to \WFH in a
software engineering context. In \cref{sec:method}, we outline our
method to include our company setting, followed by our results in
\cref{sec:results}. We conclude the paper with a discussion and a distilled set of recommendations in \cref{sec:discussion}, and final remarks in
\cref{sec:conclusion}.

\section{Background}\label{sec:background}
To provide a context for this paper, we look to the literature
on developing software during the COVID-19 pandemic, and how those working remotely are
managing to retain a level of physical and mental health. 

Several surveys
have been conducted to assess the impact of COVID-19 pandemic on
software developers, ranging from measuring practitioner wellness
\cite{Ralph_2020_Pandemic}, productivity \cite{Bao_2020_How}, and job
satisfaction and work-life balance \cite{Bellmann_2020_Working}.  Some
shift the focus from problem to solution, proposing mitigation
strategies \cite{Jones_2020_Optimal}; others~\cite{Ford_2020_Tale},
recognize the dichotomy where developers prefer, and at the same time
dislike, some aspects of working from home.

Ralph et al's \cite{Ralph_2020_Pandemic} timely and large global study
on the impact of COVID-19 on the wellness of developers (with over 2,200
responses across 53 countries), identified that poor disaster
preparedness, fear relating to the pandemic, and not having the right
set up at home (home office ergonomics), all adversely affect
well-being.  They were also, through statistical tests, able to show a
close relationship between wellness and productivity.  Other areas of
concern are indications that ``women, parents and people with disabilities
may be disproportionately affected.''  They conclude that support
needs to be tailored to the needs of individuals.

Ford et al \cite{Ford_2020_Tale} noted the dichotomy of the same
factors having both a negative and positive effect.  For example,
people missed social interactions, found it hard to create a clear
boundary between home life and work, suffered from poor ergonomics,
had less visibility and awareness of how other people are working, and
exercised less.  Communication was also an issue, and some parent employees
suffered from a lack of childcare.

On the other hand, the same participants see benefits to working from home,
such as, no commute and associated reduction in expense, flexible
hours, being close to family, comfort at home, health, and more time.
This dichotomy was also observed by Boa et al \cite{Bao_2020_How},
``Many \ldots agreed that [\WFH] can have both positive and negative
impacts on developer productivity.''

A big question that will guide the future of \WFH post pandemic is
whether productivity is impacted. Boa et al's  \cite{Bao_2020_How} study on the productivity before and during COVID-19
found very little difference (using a selection of productivity measures on the output of 139 developers). The literature presents mixed messages here, as these neutral findings contrast with those
of Ralph et al \cite{Ralph_2020_Pandemic} who suggest that conditions imposed by the pandemic and \WFH had a negative impact
on productivity. The other extreme comes from pre-pandemic studies on teleworking, and home
working, where several studies revealed
productivity improvements when \WFH~\cite{Bao_2020_How}. 

Moving to the open-source community, a special report published by the GitHub Data Science team \cite{Forsgren_2020_Octoverse} highlights trends and insights into developer activity on GitHub at the start of COVID-19. The research team investigated how a sudden shift to working from home affected developers according to three themes: productivity and activity, work cadence, and collaboration. 
Their findings suggest that developers have continued to contribute and show resilience in the face of uncertainty. 
Developers are working longer, by up to an hour a day. ``The cadence
of work has changed.~\cite[\emph{Key Findings}]{Forsgren_2020_Octoverse}'' The researchers suggest that these longer workdays (when working from home) may be due to non-work interruptions such as childcare, that cannot be ignored when working from home.
The team warn that ``patterns of developer activity have implications for burnout'' \cite{Forsgren_2020_Octoverse}. 

Transitioning to new work routines can lead to developers spending more time online, and completing tasks on time might be taking away from ``personal time and breaks to replenish, ponder, and maintain healthy separation.~\cite[\emph{Key Findings}]{Forsgren_2020_Octoverse}''  They query whether this is sustainable.  On a positive side, developers are collaborating more, with many open-source projects seeing a spike in activity.


Key cross cutting themes from the research on development during the COVID-19 pandemic
are: maintaining a work-life balance with boundaries to avoid burnout, setting up home
office environment, childcare, women being disadvantaged due to care
responsibilities, productivity changes, lack of awareness of others' work,
communication, and exercise.

The next section looks at how we in \Ocuco responded and implemented changes to support our employees during the pandemic.

\section{Method}\label{sec:method}
\begin{figure*}
  \centering{
    \includegraphics[width=2.0\columnwidth]{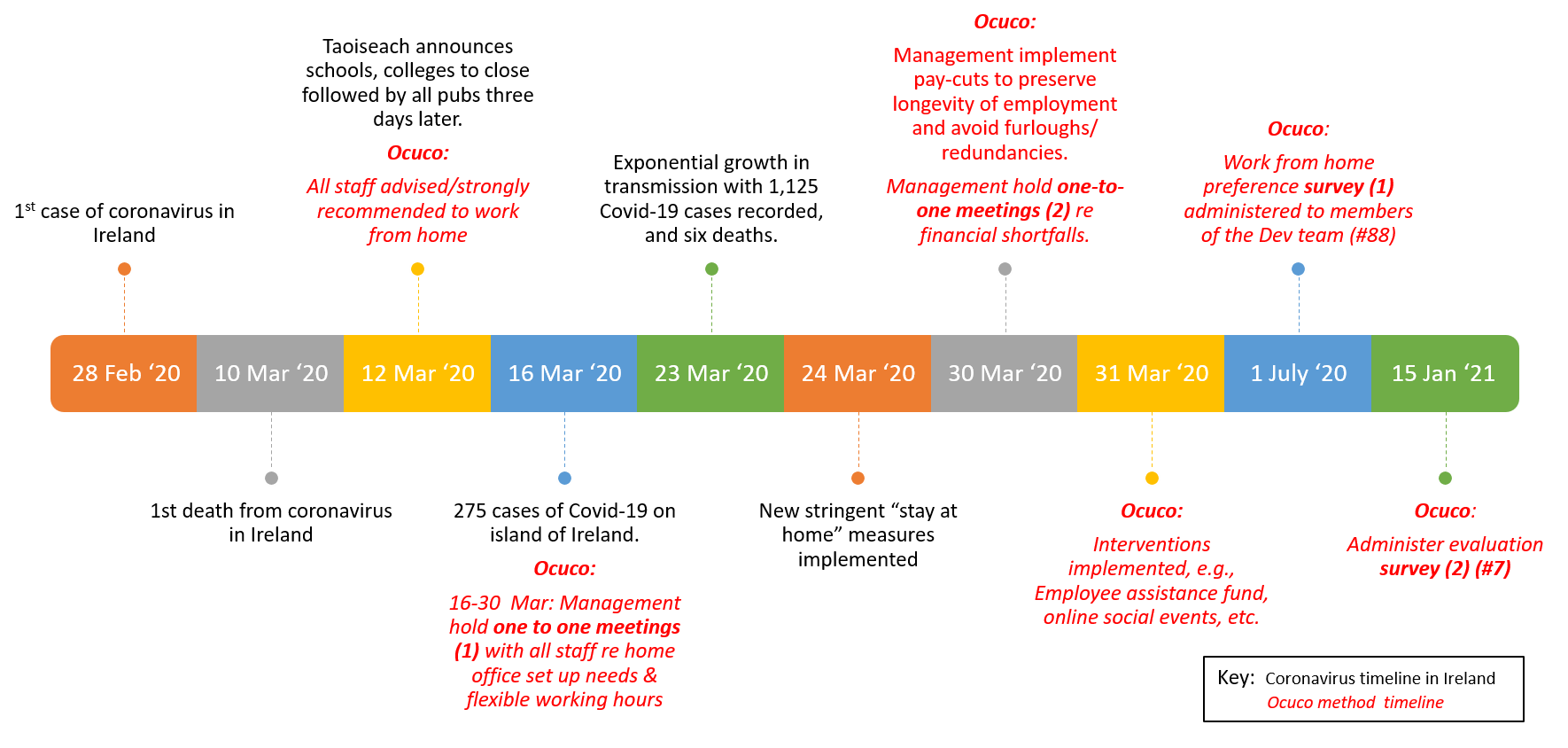}}
  \caption{Coronavirus Timeline in Ireland and \Ocuco's response}
  \label{fig:ocuco-pandemic-timeline}
\end{figure*}

To answer our question, \RQ  we looked
at wellness according to three key concepts: \emph{People},
\emph{Technology}, and
\emph{Process}. \cref{fig:ocuco-pandemic-timeline} shows the timeline
of \Ocuco interventions in response to the emergence of COVID-19 in Ireland.

\subsection{\Ocuco Setting}\label{sec:ocuco-setting}

\begin{figure*}
  \centering{
    \includegraphics[width=2.0\columnwidth]{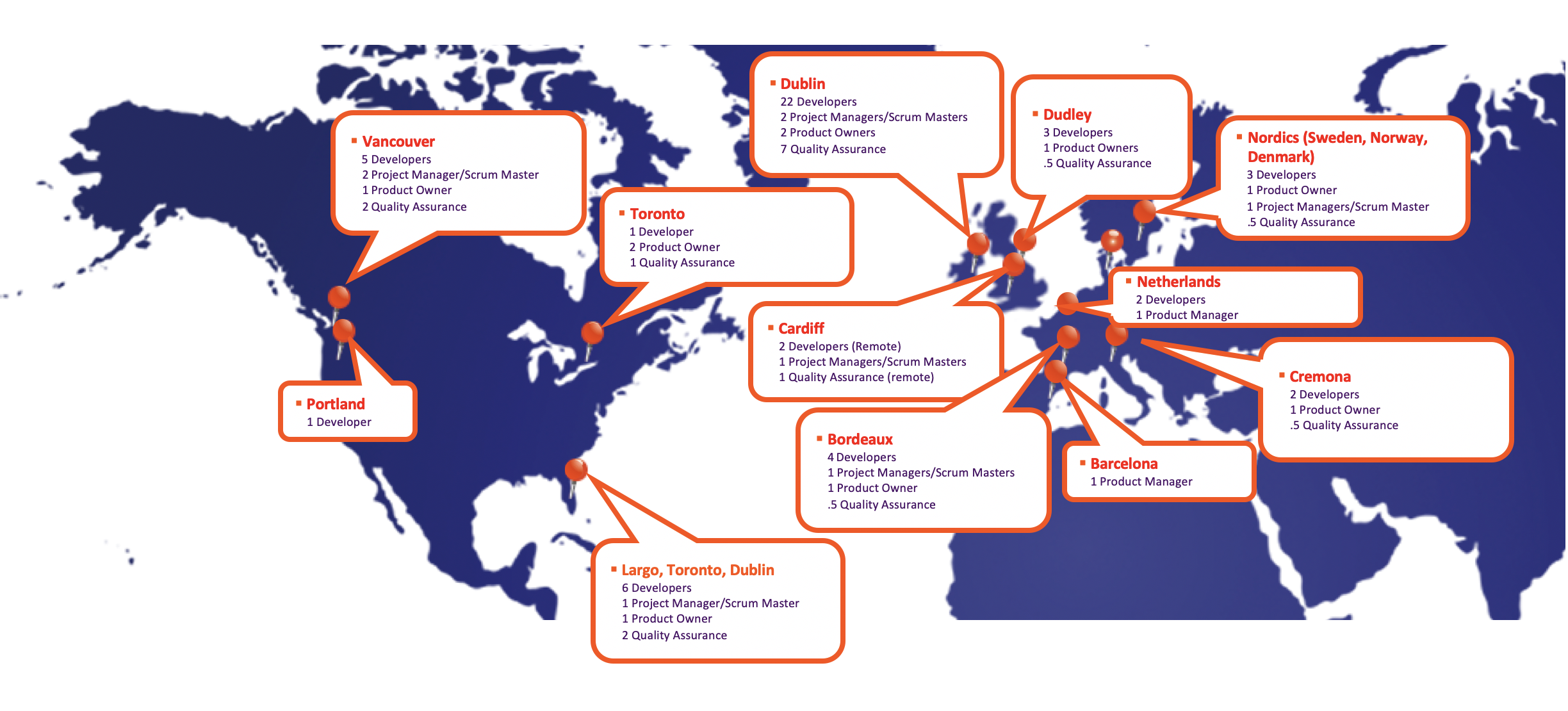}}
  \caption{\Ocuco.}
  \label{fig:ocuco-limited}
\end{figure*}

Our company, \Ocuco, is a medium-sized Irish software company that
develops practice and laboratory management software for the optical
industry. We employ more than 300 staff members in our software development organization (including support and management personnel). Of these, a growing team of 75 developers and 40 operations engineers work from \Ocuco's Dublin Headquarters, working on software development projects across twelve countries.  \cref{fig:ocuco-limited} shows the distribution of countries and roles.

\subsection{Problem identification, data collection and analysis}

We conducted a series of one-to-one meetings in March 2020. The first
set of meetings focused on checking staff welfare and needs in
transitioning to \WFH; the second set of meetings were held straight
after the announcement of pay cuts, in which financial assistance was
offered where necessary. 
A ``work location'' survey was administered to 88 employees in July 2020 (to include developer, product owner, QA and project manager roles) based across all geographic locations. With many members of \Ocuco already working remotely, we wanted a picture of changes imposed by the \WFH regime, and future wishes. The survey asked participants to select from one of seven options relating to \WFH and/or working from the office. 
The data analysis involved identifying individual needs (from the one to one meetings), and aggregating the responses from future work location survey. 
Interventions based on this analysis were implemented with immediate effect in March, 2020 as described in
\cref{sec:results}.

\subsection{Evaluation of impact of changes made during the pandemic}

In order to check management perception directly as to how well the
interventions were working, we administered an evaluation survey on
15th January 2021 to a
stratified sample,  to assess People, Technology,
Process, and Wellness factors  (see \cref{sec:survey}). The online survey was administered using
Microsoft Forms\footnote{\url{forms.office.com}.} in which all responses were anonymous and voluntary.

The sample comprised ten team members having
one of seven roles: Senior Software
Engineer (x 2), Software Engineer (x 3), QA Manager, QA Engineer,
Automated Test Architect, Project Manager, Development \& PMO
Director;  we received seven responses.   Because some of these roles
have only one team member in the sample, to preserve anonymity the
survey does not ask respondents to identify their roles; nevertheless,
we can deduce that at least four of the seven roles are represented in
our results (\cref{sec:results}).

\section{ Results}\label{sec:results}

This section presents results derived from our data collection, problem identification and interventions according to the timeline given in \cref{fig:ocuco-pandemic-timeline}. First we present results from our `work location preference' survey in \cref{tab:work_location_pref},
followed by a list of problems and interventions 
to support employees \WFH (WFH) in \cref{tab:new-strategies-people,tab:new-strategies-process,tab:new-strategies-technology,tab:new-strategies-wellness}.  Finally, \cref{tab:survey-results} presents preliminary results from our \WFH evaluation survey that we administered in January 2021 (see \cref{sec:survey} for questions). 
\cref{tab:survey-demographics} provides the demographic breakdown of the seven practitioners who responded to our survey.  
Note, we cannot specify roles of respondents  as we did not include this identifying feature in our preliminary survey, to preserve anonymity.
However, we can be confident that we have at least four roles included in the responses (see breakdown of roles in \cref{sec:method}). 
\begin{table}
 \caption{Work location preference (\#88) - Home/Office/Abroad?} \label{tab:work_location_pref}
{\small
  \begin{tabular}{lcc}
    \toprule
    Work location (pre COVID-19 and future)                 & \multicolumn{2}{c}{Count} \\
    \midrule
    No change, worked from abroad before:                   & 18 & 20\%                 \\
    No change, \WFH (WFH) full time before:                 & 2  & 2\%                  \\
    Need to be in office full time (once conditions allow): & 4  & 5\%                  \\
    Change: Would like flexibility (50 WFH/50 office):      & 32 & 36\%                 \\
    Change: Would like flexibility - mainly WFH:            & 14 & 16\%                 \\
    Change: Would like flexibility -  mainly office:        & 13 & 15\%                 \\
    Change: Would like flexibility - other country/city:    & 5  & 6\%                  \\
    \bottomrule
  \end{tabular}
}
\end{table}

Although all Irish based staff were asked to \workFromHome to keep safe and comply with Ireland's COVID-19 guidance, results from our world-wide work location survey (\cref{tab:work_location_pref}) shows 88 practitioners (comprising developers, product owners, QAs and project managers) had a mix of preferences.
20\% already \workFromHome (prior to COVID-19). 59\% (32 +14+13) of
office-based workers would like flexibility to \workFromHome or
office. 8\% of office-based workers would like to \workFromHome or
another country full-time (with willingness to go to office/Dublin
whenever required, e.g. for PI planning meetings).  A small number (5\%) feel the need to work in office full time in the future.  The majority wanted the flexibility to work from both the office and home.

\subsection{Problem identification}
The recurrent one-to-one online meetings identified anxieties, needs and preferences, as follows:

\subsubsection{People concerns}

\begin{asparaenum}
\item How to share work with childcare?

\item How to relieve financial demands such as mortgage payments, due to reduction in salary or
partner's loss of job?

\item How to facilitate move from office to \WFH, and compensate for reduced level of social and informal interaction with colleagues.

\item How to welcome new staff remotely?

\item How to support (new) staff moving to Ireland from abroad?
\end{asparaenum}

\subsubsection{Technology concerns}

\begin{asparaenum}
\item Is our current technology adequate for us to adapt to working from home, and does it scale?

\item Does everyone have the right set up at home?

\item Does need for fast broadband connection from home result in any additional cost to our staff?
\end{asparaenum}

\subsubsection{Process concerns}

\begin{asparaenum}
\item How can we adapt processes to keep people socially connected
  (beyond having the right technology)? For example, pre-pandemic, we
  held regular face-to-face program increment (PI) planning meetings
  that take place over two weeks at the Dublin headquarters.
  Attendees from 10 different countries in five time zones all travel
  to be physically in the same space.  During the pandemic we held a
  virtual version of the PI planning meeting, and in the retrospective
  attendees reported that they ``miss social contact and cross team
  collaboration enabled by on-site meetings.''

\item How can we replicate (in a virtual setting) the informal gatherings and interactions enjoyed by our staff?  
While ``daily stand-ups'' and other \SAFelong (\SAFe) virtual ceremonies offer opportunities for staff to meet, many employees noted that this did not satisfy the need for informal gatherings, and getting to know new members in a relaxed social setting.

\end{asparaenum}

\subsection{Interventions}

We introduced new initiatives in recognition of the needs expressed by employees in our one-to-one interviews. 

\cref{tab:new-strategies-people,tab:new-strategies-process,tab:new-strategies-technology} list our interventions to address the above problems.
All interventions were implemented immediately, and where appropriate,
repeated regularly.

\begin{table}
\caption{New strategies--People}\label{tab:new-strategies-people}

\begin{tabular}{p{.4\columnwidth}p{.55\columnwidth}}
\toprule
Problem                                                                                   & Solution \tabularnewline
\midrule
Sharing work with childcare.                                                              
& Allowed flexibility to staff with kids who are \WFH--could adjust their day to share childcare. \tabularnewline
\midrule
Unsuitable home environment for remote working (e.g. no space, too many people in house). 
& Offered a Taxi service to office (adapted to be COVID-19 compliant) to those unable to \WFH comfortably. \tabularnewline
\midrule
Concerns about family who are living in a different country. 
& Facilitated staff living abroad to go home for extended periods to see their families allowing for COVID-19 restrictions. \tabularnewline
\midrule
Concerns about living arrangements for those committed to moving countries. 
& Provided accommodation to staff moving to Ireland to start new roles within \Ocuco. \tabularnewline
\midrule
Loss of contact.                                                                           
& Line Managers check in with teams at least once a month. (Teams meet regularly with their daily stand-ups). \tabularnewline
\bottomrule
\end{tabular}
\end{table}

\begin{table}
\caption{New strategies -- Process}\label{tab:new-strategies-process}

\begin{tabular}{p{.4\columnwidth}p{.55\columnwidth}}
\toprule

Problem & Solution\tabularnewline
\midrule
Feeling isolated. 
& Established `Remote working Initiatives' Team to keep people connected. Simulated team lunches/nights out, coffee dock/water cooler chats, etc.  Required everyone in teleconferences to share their cameras.  \tabularnewline
\midrule
Financial difficulties.
& Introduced Employee Assistance fund for those struggling during temporary pay cut period. \tabularnewline
\midrule
Lack of job security.
& Offered free and independent advice to staff with financial commitments, e.g. whether to avail of mortgage extensions offered by their banks. \tabularnewline
\midrule
Anxiety over status of \Ocuco business.
& CEO gave key updates at least once a month in company wide meetings. \tabularnewline
\bottomrule
\end{tabular}
\end{table}

\begin{table}
\caption{New strategies -- Technology}\label{tab:new-strategies-technology}

\begin{tabular}{p{.4\columnwidth}p{.55\columnwidth}}
\toprule
Problem & Solution\tabularnewline
\midrule
Home office and ergonomics.
& Reached out to ensure all staff had a suitable \WFH environment; budget provided for chairs, desks, and extra screens. \tabularnewline
\midrule
How to scale existing video conferencing applications.
& Switched to Microsoft Teams\textregistered. \tabularnewline
\midrule
Retain privacy and security of code and data.
& Provided access to virtual build machines via virtual private network (VPN). \tabularnewline
\midrule
Connection costs.
& Paid broadband subsidies and upgrade costs where needed. \tabularnewline
\bottomrule
\end{tabular}
\end{table}

\begin{table}
\caption{New strategies -- Wellness}\label{tab:new-strategies-wellness}

\begin{tabular}{p{.4\columnwidth}p{.55\columnwidth}}
\toprule
Problem & Solution\tabularnewline
\midrule
Anxiety and fear of burn-out.
& Check for burn-out, and encourage setting work/home boundaries. Meet staff one-to-one meetings, or administer anonymous surveys in work routines and levels of anxiety can be reported anonymously or directly with line manager. \tabularnewline
\midrule
New employee feelings of uncertainty.
& Set-up remote staff induction programs to ensure new staff are on-boarded efficiently and quickly to the \Ocuco team. Provide accommodation to staff committed to moving to Ireland to start new roles. \tabularnewline
\midrule
All work and no play.
& We maintain social connections globally by hosting: Pizza Fridays (with synchronized expensed lunches), knowledge and music quizzes (\cref{fig:zoom-social-mtg}). Coffee mornings where groups of four from different countries and offices meet via video conference for casual chat. ``Coffee Dock~\footnote{The ``coffee dock'' is a self-service espresso bar located in \Ocuco{}'s physical headquarters.}'' meet ups where anyone can open up a `meet' an invite colleagues. \tabularnewline
\midrule
Feeling lost and isolated.
& The management team has a monthly informal one-to-one chat with each staff member. \tabularnewline
\bottomrule
\end{tabular}
\end{table}

\subsection{Evaluation Survey}
\begin{table}[t]
\caption{Survey (\cref{sec:survey}) respondent demographics (\#7) (`DNS' = ``did not specify'').} \label{tab:survey-demographics}
{\small
  \centering
  \begin{tabular}{lcccc}
    \toprule
    Category                                 & \multicolumn{4}{c}{Number}     \\
    \midrule
    \multirow{2}{*}{Gender:}                 & Female & Male    & Other & DNS \\
                                             & 3      & 4       & -     & -   \\
    \midrule                                 
    \multirow{2}{*}{Age:}                    & 20-29  & 30-39   & 40-49 & DNS \\
                                             & 1      & 2       & 3     & 1   \\
    \midrule                                 
    \multirow{2}{*}{Location:}               & GB     & Ireland & DNS   &     \\
                                             & 1      & 5       & 1     &     \\
    \midrule
    \multirow{2}{*}{Time at \Ocuco (years):} & 1-2    & 3-5     & 5-10  & 10+ \\
                                             & 1      & 1       & 1     & 4   \\
    \bottomrule
  \end{tabular}
}
\end{table}

On 15 January, 2021, we administered an online ``Work from home evaluation'' survey (see \cref{sec:survey} for questions).
We received \numberofresponses completed responses (out of 10);
\cref{tab:survey-demographics}  shows the sample
distribution. Information has been aggregated to ensure anonymity in
this small sample.  

\begin{figure}[t]
  \centering%
  \hspace*{-.05\columnwidth}\includegraphics[width=1.01\columnwidth]{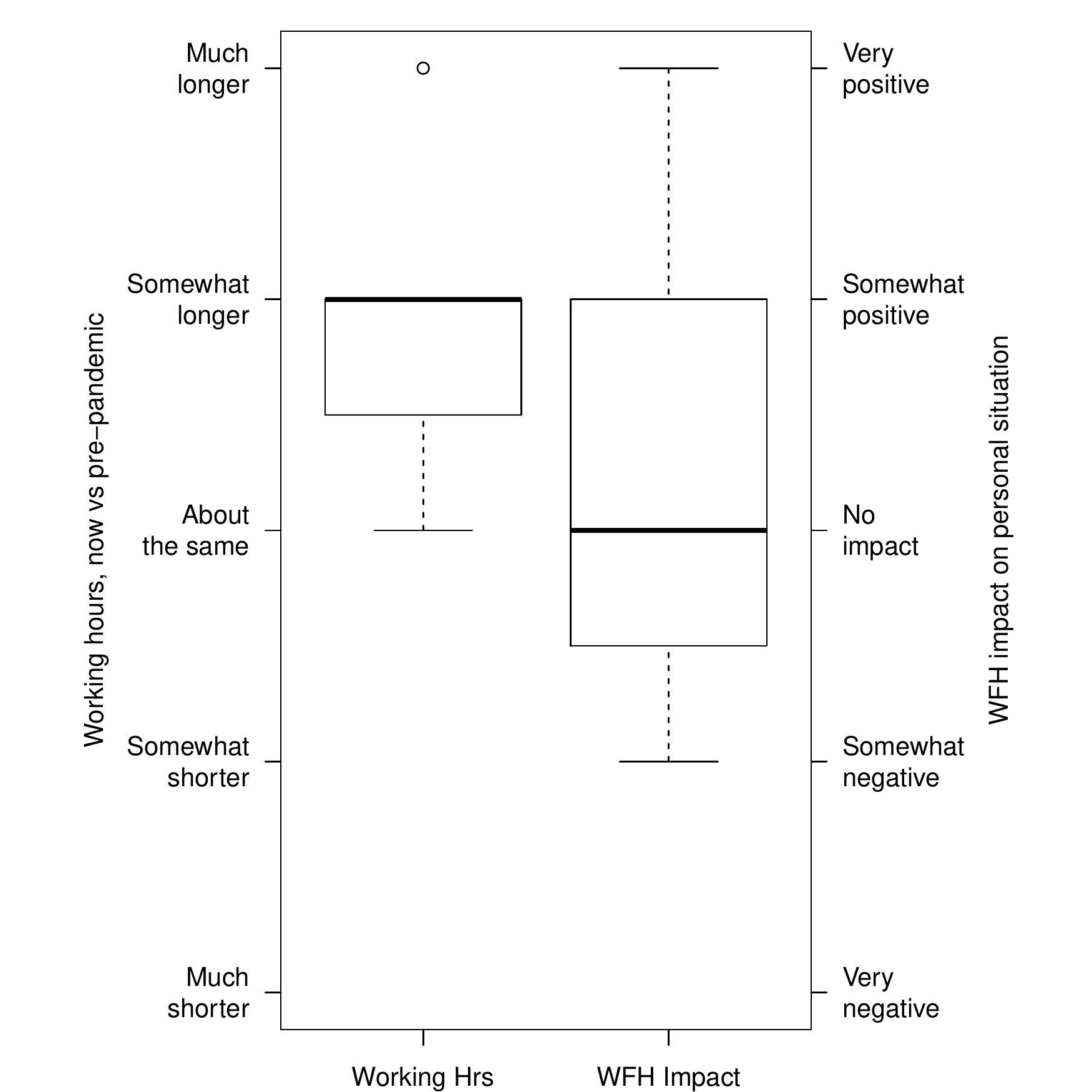}
  \caption{Plot of survey results re working hours (Q1.7, \cref{sec:survey})  and impact
    on personal life (Q2.7) (\#7).}  \label{fig:survey-results}
  \Description{Two box-plots summarizing the survey results in
    \cref{tab:survey-results} related to
    working hours and impact on personal life.}
\end{figure}

\begin{figure}[t]
  \centering%
  \hspace*{-.0\columnwidth}\includegraphics[width=1.01\columnwidth]{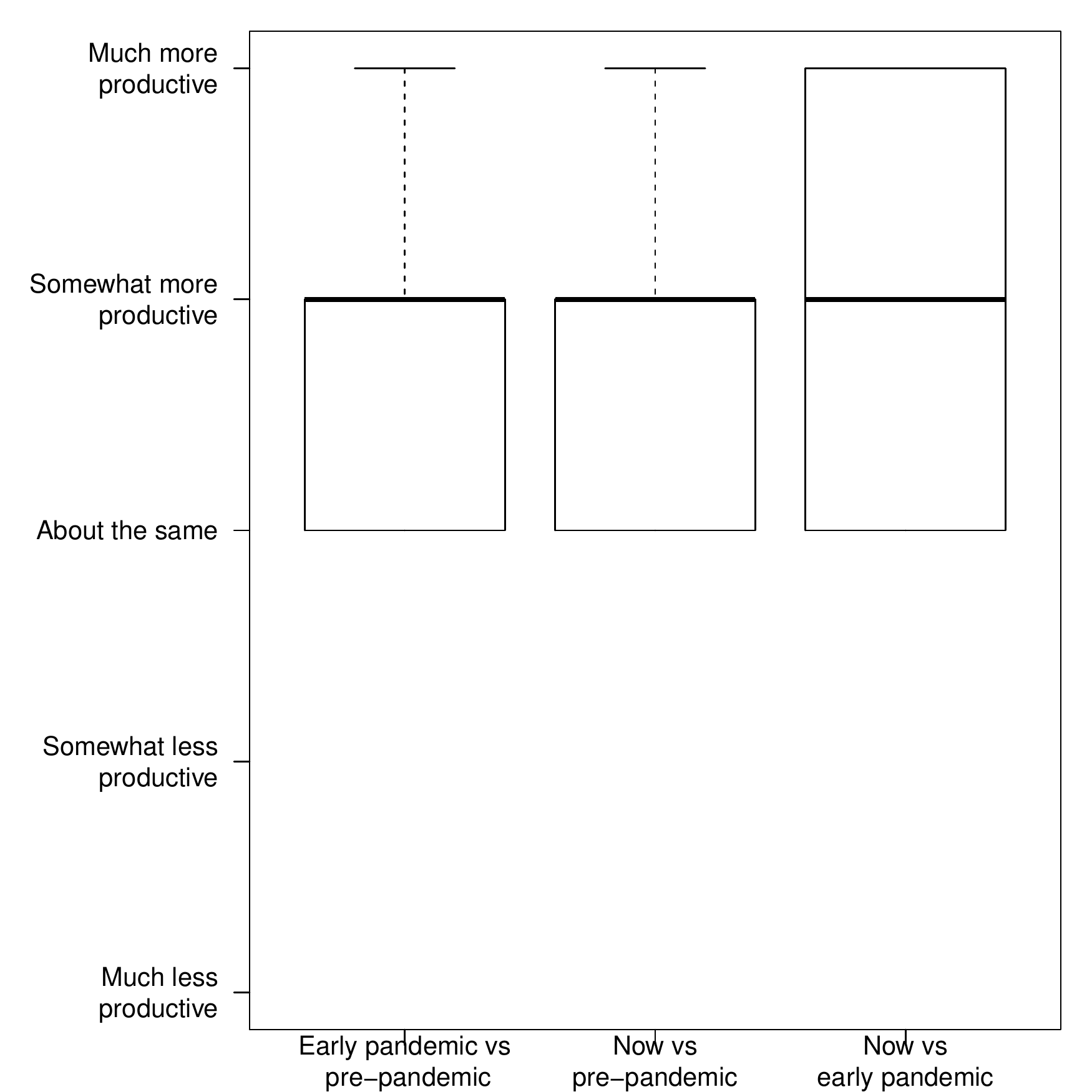}
  \caption{Plot of survey results related to productivity (Q4.5-7, \cref{sec:survey}) (\#7).}  \label{fig:productivity-results}
  \Description{Three box-plots summarizing the survey results  in
    \cref{tab:survey-results} related to productivity.}
\end{figure}

A selected set of Evaluation Survey results are shown in
\cref{tab:survey-results}, \cref{fig:survey-results} and \cref{fig:productivity-results}.  Even with our small sample, we can draw some conclusions from
these results with a degree of confidence
\footnote{According to the ``rule of five [participants]'' \cite{Hubbard_2014_How} there is nearly a 95\% likelihood that the median response of
  a population is within the highest and lowest responses of a random
  sample of only five members of the population. So  with a
  sample of seven we have high confidence that the median response
  represents the company as a whole.} Looking at the level of support
  from Ocuco (Q2.1, \cref{tab:survey-results}), there was near consensus that the support was excellent. Given the range of responses for other questions, we describe  further in \cref{fig:survey-results} and \cref{fig:productivity-results}.

\cref{fig:survey-results}
suggests that \Ocuco employees are working at least as
long when \WFH as when they were in the office.
\cref{fig:survey-results} also shows that \Ocuco developers perceive they are at least as
productive during the pandemic as they were before the pandemic.  There was a more mixed response when
considering the impact of \WFH on personal responsibilities.

Regarding productivity, we wanted to know whether there was a change
in productivity at the beginning of the pandemic, when \WFH was novel,
and little was known about short- or long-term effects of
the pandemic.  The responses suggest increased productivity
immediately after shifting to \WFH.  This increase in productivity
seems to have been maintained, with a majority of respondents continuing to
report higher productivity compared to the ``pre-pandemic''
(\cref{fig:productivity-results}).  
Finally, looking at the evolution from when we first
moved to \WFH to now, we see a further upward shift in productivity.  This could
be due to taking time to acclimatize to the early move to \WFH.
But we don't know whether this productivity increase is the result of longer
working hours, or other factors related to \WFH specifically.  Also,
these observations are based on participant's self-reported
perceptions; it remains to be seen whether actual productivity has
increased.

\begin{table*}
  \caption{Selected results from \Ocuco's COVID-19 \WFH evaluation survey (\#7) (see \cref{sec:survey}).} \label{tab:survey-results}
  {\small
    \centering%
    \begin{tabular}{L{.22\textwidth}C{.13\textwidth}C{.13\textwidth}C{.13\textwidth}C{.13\textwidth}C{.13\textwidth}}
      \toprule
      \textbf{Working hours now vs pre-pandemic (Q1.7)} & Much longer & Somewhat longer & About the same & Somewhat shorter & Much shorter\\
\midrule
 & 1 & 4 & 2 & 0 & 0 \\ 
\midrule
\textbf{Level of support from Ocuco (Q2.1)} & Excellent & Adequate & Neither helped nor hindered & Inadequate & Very inadequate \\ 
\midrule
 & 6 & 1 & 0 & 0 & 0 \\ 
\midrule
\textbf{\Wfh impact on personal responsibilities (Q2.7)} & Very positive & Somewhat positive & No impact & Somewhat negative & Very negative\\
\midrule
 & 2 & 0 & 3 & 2 & 0 \\ 
\midrule
\textbf{Productivity} & Much more productive & Somewhat more productive & About the same & Somewhat less productive & Much less productive\\
\midrule
{Early- vs pre-pandemic (Q4.5)} & 1 & 3 & 3 & 0 & 0 \\ 
{Now vs pre-pandemic (Q4.6)} & 1 & 3 & 3 & 0 & 0 \\ 
{Now vs early--pandemic (Q4.7)} & 3 & 1 & 3 & 0 & 0 \\ 
 
      \bottomrule
    \end{tabular}
  }
\end{table*}

\section{Discussion}\label{sec:discussion}

Politicians, organizations, economists and individuals are all consumed with the pressing economic question of the
long-term cost arising from the COVID-19 pandemic~\cite{Gottlieb_2020_Working}.  Organizations need to balance keeping afloat and surviving during the pandemic, maintaining an experienced and trained workforce, and, being prepared for growth in the near future.  Agile takes on a whole new meaning in the pandemic, moving way beyond the development team. SAFe paves the way for the whole company to be involved, so that decisions can be made quickly, communication channels are open with regular contact across roles and divisions, and management are flexible in terms of changing processes to meet the new needs of employees.  As a people-intensive field,
software engineering relies heavily on the wellness of a skilled
workforce. A particular challenge is to reverse the negative impact \WFH can have on work-life balance~\cite{Bellmann_2020_Working}. From our initial survey feedback, we need to be particularly aware of raised
anxiety in our employees and longer working hours.

\subsection{Recommendations}

Returning to our original research question, \RQ we distill the
results from the previous section into five recommendations:

\paragraph{Recommendation 1: Be flexible.}

Example: allow flexible working hours, and flexibility of work location.
Allow employees who have relocated from abroad to return to their home
country, when travel restrictions allow.

\paragraph{Recommendation 2: Be proactive and supportive.}

Example:  reach out to all staff to ensure they have a suitable work
from home set up.  Provide financial support for acquiring same.
Facilitate staff living away from their families to either find a place to live (for new members), or travel home when COVID-19 restrictions allow.

\paragraph{Recommendation 3: Keep connected.}

Example: Remove technical barriers to communication.  Initiate and
maintain communication.  Keep lines of communication open, bi-directional, and active.  Balance number
of work meetings with regular and varied social events to on-board new
members of staff and keep existing people connected.  Have fun!

Due to the wide distribution of sites, \Ocuco had already invested in
strong tools to collaborate and host meetings across distance.
As such, technology interventions during COVID-19 were more
about scaling, upgrading personal connections,  and ensuring privacy, than introducing new applications. 

We are planning more remote events during 2021 to replace the larger ``all hands''
workshops that normally take place in person.  In addition, we are 
coordinating the safe return to office day events when
local government guidelines permit.

\paragraph{Recommendation 4: Give financial support where needed.}

Allocate funds to support staff experiencing financial shortfalls
as a result of temporary pay cuts.
Offer impartial, professional independent advice on financial affairs,
such as bank loans and mortgages.

\paragraph{Recommendation 5: Show strong leadership}

Have the CEO and Management team provide regular updates on the current and future direction of the company, and provide an open door policy for staff to come and ask questions or raise concerns.

\Ocuco is a medium-size company, and so can implement
recommendations such as these rapidly, on a company-wide basis.  Nevertheless,
we feel that larger organizations can still benefit from these
recommendations, which can be adapted to the department or
division level.  Not doing so, on the other hand, risks depletion of
the social capital created from face-to-face working~\cite{Franklin_2020_Remote}.

\begin{figure}
\includegraphics[width=0.95\columnwidth]{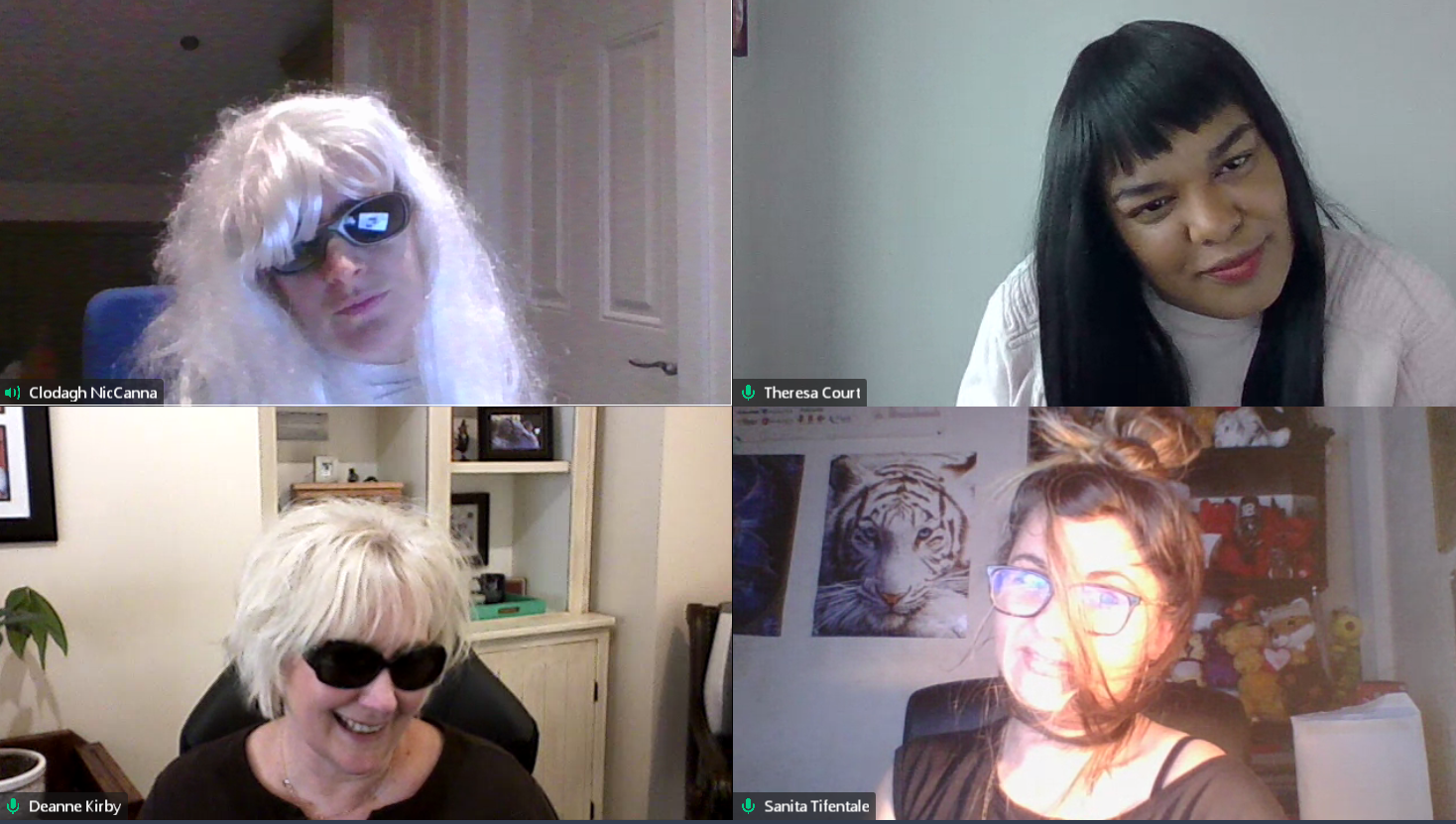}

\caption{Virtual music quiz hosted by \Ocuco's BeNeLux team.}\label{fig:zoom-social-mtg}
\Description{Screen capture of a virtual movie quiz conducted via
  video conference, showing four participants wearing wigs and
  sunglasses to resemble rock musicians.}
\end{figure}

\section{Conclusion}\label{sec:conclusion}

In this paper we presented our experience of managing remote development
teams during the COVID-19 pandemic.  \Ocuco has taken very naturally to remote working, having already
had a strong ethos, based on agile practices, for working with
colleagues in any of the 14 different \Ocuco locations around the
world.  
But we recognized early
that our previous processes and infrastructure do not guarantee a smooth transition to
\WFH full time. As a result, we implemented several new
interventions, in three key areas: People, Process and
Technology. Our key message when it comes to \emph {People} is that management need to be quick to recognize and react to the crisis, show strong leadership, listen to employees' needs, and be informative. \emph{
Technology}, on the other hand, as the route to remaining connected,
must be ubiquitous throughout the company, regardless of
location; equipment and infrastructure costs must be met by the
employer.  Finally, \emph{Process} changes need to be fast, flexible, and inclusive.

Our results are encouraging, as during the pandemic crisis, we have kept all our current staff on the payroll, employed new people, and won new contracts. Early feedback on productivity is also positive. 

We still have a lot of hard work to do to prevent losing the social
fabric of the company and the culture we have worked so hard to build over 25 years of trading.
So, although we've been successful in working remotely in the past, we recognize
there is more to do, and we do all miss the office for the human
interactions.

Our future plans involve administering the \Ocuco COVID-19 Working from Home Evaluation Survey to all \Ocuco employees.

\section{Acknowledgments}

We are indebted to the
many members of \Ocuco who responded to our survey and volunteered
their insights about working during the COVID-19 pandemic. This work was supported, in part, by Science Foundation Ireland grant
13/RC/2094 to Lero -- The SFI Software Research Centre.

\bibliographystyle{ACM-Reference-Format}
\interlinepenalty=10000         
\bibliography{ms}


\begin{thebibliography}{10}


\ifx \showCODEN    \undefined \def \showCODEN     #1{\unskip}     \fi
\ifx \showDOI      \undefined \def \showDOI       #1{#1}\fi
\ifx \showISBNx    \undefined \def \showISBNx     #1{\unskip}     \fi
\ifx \showISBNxiii \undefined \def \showISBNxiii  #1{\unskip}     \fi
\ifx \showISSN     \undefined \def \showISSN      #1{\unskip}     \fi
\ifx \showLCCN     \undefined \def \showLCCN      #1{\unskip}     \fi
\ifx \shownote     \undefined \def \shownote      #1{#1}          \fi
\ifx \showarticletitle \undefined \def \showarticletitle #1{#1}   \fi
\ifx \showURL      \undefined \def \showURL       {\relax}        \fi
\providecommand\bibfield[2]{#2}
\providecommand\bibinfo[2]{#2}
\providecommand\natexlab[1]{#1}
\providecommand\showeprint[2][]{arXiv:#2}

\bibitem[\protect\citeauthoryear{Bao, Li, Xia, Zhu, Li, and Yang}{Bao
  et~al\mbox{.}}{2020}]%
        {Bao_2020_How}
\bibfield{author}{\bibinfo{person}{Lingfeng Bao}, \bibinfo{person}{Tao Li},
  \bibinfo{person}{Xin Xia}, \bibinfo{person}{Kaiyu Zhu}, \bibinfo{person}{Hui
  Li}, {and} \bibinfo{person}{Xiaohu Yang}.} \bibinfo{year}{2020}\natexlab{}.
\newblock \showarticletitle{How does Working from Home Affect Developer
  Productivity?--A Case Study of Baidu During COVID-19 Pandemic}.
\newblock \bibinfo{journal}{\emph{arXiv preprint arXiv:2005.13167}}
  (\bibinfo{year}{2020}).
\newblock


\bibitem[\protect\citeauthoryear{Bellmann and H{\"u}bler}{Bellmann and
  H{\"u}bler}{2020}]%
        {Bellmann_2020_Working}
\bibfield{author}{\bibinfo{person}{Lutz Bellmann} {and} \bibinfo{person}{Olaf
  H{\"u}bler}.} \bibinfo{year}{2020}\natexlab{}.
\newblock \showarticletitle{Working from home, job satisfaction and work--life
  balance--robust or heterogeneous links?}
\newblock \bibinfo{journal}{\emph{International Journal of Manpower}}
  (\bibinfo{year}{2020}).
\newblock


\bibitem[\protect\citeauthoryear{Ford, Storey, Zimmermann, Bird, Jaffe,
  Maddila, Butler, Houck, and Nagappan}{Ford et~al\mbox{.}}{2020}]%
        {Ford_2020_Tale}
\bibfield{author}{\bibinfo{person}{Denae Ford}, \bibinfo{person}{Margaret-Anne
  Storey}, \bibinfo{person}{Thomas Zimmermann}, \bibinfo{person}{Christian
  Bird}, \bibinfo{person}{Sonia Jaffe}, \bibinfo{person}{Chandra Maddila},
  \bibinfo{person}{Jenna~L Butler}, \bibinfo{person}{Brian Houck}, {and}
  \bibinfo{person}{Nachiappan Nagappan}.} \bibinfo{year}{2020}\natexlab{}.
\newblock \showarticletitle{A tale of two cities: Software developers working
  from home during the covid-19 pandemic}.
\newblock \bibinfo{journal}{\emph{arXiv preprint arXiv:2008.11147}}
  (\bibinfo{year}{2020}).
\newblock


\bibitem[\protect\citeauthoryear{Forsgren}{Forsgren}{2020}]%
        {Forsgren_2020_Octoverse}
\bibfield{author}{\bibinfo{person}{N. Forsgren}.}
  \bibinfo{year}{2020}\natexlab{}.
\newblock \bibinfo{title}{Octoverse spotlight: An analysis of developer
  productivity, work cadence, and collaboration in the early days of covid-19}.
\newblock \bibinfo{howpublished}{WWW page, accessed 10 January 2021}.
\newblock
\urldef\tempurl%
\url{https://github.blog/2020-05-06-octoverse-spotlight-an-analysis-of-developer-productivity-work-cadence-and-collaboration-in-the-early-days-of-covid-19/}
\showURL{%
\tempurl}


\bibitem[\protect\citeauthoryear{Franklin}{Franklin}{2020}]%
        {Franklin_2020_Remote}
\bibfield{author}{\bibinfo{person}{Neil Franklin}.}
  \bibinfo{year}{2020}\natexlab{}.
\newblock \bibinfo{title}{Remote working productivity will slump as firms burn
  up their social capital}.
\newblock \bibinfo{howpublished}{{\emph{insight}} online magazine, accessed 6
  March 2021}.
\newblock
\urldef\tempurl%
\url{workplaceinsight.net/remote-working-productivity-will-slump-as-firms-burn-up-their-social-capital}
\showURL{%
\tempurl}


\bibitem[\protect\citeauthoryear{Gottlieb, Grobov{\v{s}}ek, and
  Poschke}{Gottlieb et~al\mbox{.}}{2020}]%
        {Gottlieb_2020_Working}
\bibfield{author}{\bibinfo{person}{Charles Gottlieb}, \bibinfo{person}{Jan
  Grobov{\v{s}}ek}, {and} \bibinfo{person}{Markus Poschke}.}
  \bibinfo{year}{2020}\natexlab{}.
\newblock \showarticletitle{Working from home across countries}.
\newblock \bibinfo{journal}{\emph{Covid Economics}} \bibinfo{volume}{1},
  \bibinfo{number}{8} (\bibinfo{year}{2020}), \bibinfo{pages}{71--91}.
\newblock


\bibitem[\protect\citeauthoryear{Hubbard}{Hubbard}{2014}]%
        {Hubbard_2014_How}
\bibfield{author}{\bibinfo{person}{Douglas~W. Hubbard}.}
  \bibinfo{year}{2014}\natexlab{}.
\newblock \bibinfo{booktitle}{\emph{How to Measure Anything: Finding the Value
  of Intangibles in Business} (\bibinfo{edition}{3} ed.)}.
\newblock \bibinfo{publisher}{Wiley}.
\newblock


\bibitem[\protect\citeauthoryear{Jones, Philippon, and Venkateswaran}{Jones
  et~al\mbox{.}}{2020}]%
        {Jones_2020_Optimal}
\bibfield{author}{\bibinfo{person}{Callum~J Jones}, \bibinfo{person}{Thomas
  Philippon}, {and} \bibinfo{person}{Venky Venkateswaran}.}
  \bibinfo{year}{2020}\natexlab{}.
\newblock \bibinfo{booktitle}{\emph{Optimal mitigation policies in a pandemic:
  Social distancing and working from home}}.
\newblock \bibinfo{type}{{T}echnical {R}eport}. \bibinfo{institution}{National
  Bureau of Economic Research}.
\newblock


\bibitem[\protect\citeauthoryear{Noll, Beecham, and Richardson}{Noll
  et~al\mbox{.}}{2010}]%
        {Noll_2010_Global}
\bibfield{author}{\bibinfo{person}{John Noll}, \bibinfo{person}{Sarah Beecham},
  {and} \bibinfo{person}{Ita Richardson}.} \bibinfo{year}{2010}\natexlab{}.
\newblock \showarticletitle{Global Software Development and Collaboration:
  Barriers and Solutions}.
\newblock \bibinfo{journal}{\emph{{ACM} Inroads}} \bibinfo{volume}{1},
  \bibinfo{number}{3} (\bibinfo{date}{September} \bibinfo{year}{2010}).
\newblock


\bibitem[\protect\citeauthoryear{Paul, Baltes, Gianisa, Torkar, Kovalenko,
  Marcos, Nicole, Yoo, Xavier, Tan, et~al\mbox{.}}{Paul et~al\mbox{.}}{2020}]%
        {Ralph_2020_Pandemic}
\bibfield{author}{\bibinfo{person}{Ralph Paul}, \bibinfo{person}{Sebastian
  Baltes}, \bibinfo{person}{Adisaputri Gianisa}, \bibinfo{person}{Richard
  Torkar}, \bibinfo{person}{Vladimir Kovalenko}, \bibinfo{person}{Kalinowski
  Marcos}, \bibinfo{person}{Novielli Nicole}, \bibinfo{person}{Shin Yoo},
  \bibinfo{person}{Devroey Xavier}, \bibinfo{person}{Xin Tan}, {et~al\mbox{.}}}
  \bibinfo{year}{2020}\natexlab{}.
\newblock \showarticletitle{Pandemic programming}.
\newblock \bibinfo{journal}{\emph{Empirical Software Engineering}}
  \bibinfo{volume}{25}, \bibinfo{number}{6} (\bibinfo{year}{2020}),
  \bibinfo{pages}{4927--4961}.
\newblock


\end{thebibliography}

\appendix

\section{\Ocuco COVID-19 Working From Home Evaluation Survey}\label{sec:survey}

This  survey was administered online to employees after 11 months of lockdown.
Respondents were informed that participation is voluntary, completely anonymous, that results would be disseminated in aggregate form only, and
they had the option of not answering any particular question.

\vspace{1ex}


\scriptsize{
\begin{asparaenum}
  \item{Demographics}\label{sec:survey-demographics}

  \begin{asparaenum}
  \item What is your age?
    \begin{inparaenum}[$\square$]  
    \item under 20
    \item 20-29
    \item 30-39
    \item 40-49
    \item 50-59
    \item 60+
    \end{inparaenum}

  \item What is your gender?
    \begin{inparaenum}[$\square$]
    \item Male
    \item Female
    \item Other
    \item Prefer not to disclose
    \end{inparaenum}

  \item Where are you located?
    \begin{inparaenum}[$\square$]
    \item Ireland \& Great Britain
    \item Nordic region
    \item Continental Europe
    \item North America
    \item South Asia
    \item East Asia
    \item Africa
    \item Australia, New Zealand, Pacific Islands
    \end{inparaenum}

  \item How long have you worked for \Ocuco?
    \begin{inparaenum}[$\square$]
    \item Less than one year
    \item 1-2 years
    \item 3-5 years
    \item 5-10 years
    \item more than 10 years
    \end{inparaenum}

  \item On average, how many days in a working week (Mon to Friday) did you
    work from home \emph{pre-pandemic}?
    \begin{inparaenum}[$\square$]
    \item 1
    \item 2
    \item 3
    \item 4
    \item 5
    \end{inparaenum}

  \item On average, how many days a week in a working week (Mon to Friday) do
    you work from home \emph{now}?
    \begin{inparaenum}[$\square$]
    \item 1
    \item 2
    \item 3
    \item 4
    \item 5
    \end{inparaenum}

  \item  How have your working \emph{hours} (average per day/week) changed
    compared to pre-pandemic?\label{question:hours}
    \begin{inparaenum}[$\square$]
    \item Much longer
    \item Somewhat longer
    \item About the same
    \item Somewhat shorter
    \item Much shorter
    \end{inparaenum}
  \end{asparaenum}

  \item{People}\label{people}

  \begin{asparaenum}
  \item  How adequate was the support your received from \Ocuco to achieve your
    goal while working from home?\label{question:support}
    \begin{inparaenum}[$\square$]
    \item Very adequate (Excellent)
    \item Adequate
    \item Neither helped nor hindered
    \item Inadequate
    \item Very inadequate
    \end{inparaenum}

  \item What support (if any) did you find particularly helpful?

  \item What support (if any) did you find lacking, that you wish you had?

  \item  How adequately do your work from home facilities (desk, chair,
    monitor, broadband, etc.) support your work?
    \begin{inparaenum}[$\square$]
    \item Very well
    \item Adequate
    \item Neither help nor hinder
    \item Inadequate
    \item Very inadequate
    \end{inparaenum}

  \item What facilities (if any) did you find particularly helpful?
  \item What facilities (if any) did you find lacking, that you wish you had?

  \item  Has working from home impacted your personal situation (for example,
    childcare or taking care of an elderly family member)?\label{question:impact}
    \begin{inparaenum}[$\square$]
    \item Very positive impact. Example: I can now do important things I
      wasn't able to do before.
    \item Somehat positive impact. Example: I can now do pleasant things, like
      spending more time with my children.
    \item No impact.
    \item Somewhat negative impact. Example: my children want to play with me
      rather than do their schoolwork.
    \item Very negative impact. Example: I have to compete with my partner
      and/or children for private space/internet bandwidth for meetings.
    \end{inparaenum}
  \item  Please add any additional comment about your working from home
    situtation you would like to make.

  \end{asparaenum}

  \item{Technology}\label{technology}

  \begin{asparaenum}
  \item  What technology do you use frequently to meet synchronously
    with the global or local team (select all that apply)?
    \begin{inparaenum}[$\square$]
    \item Microsoft Teams
    \item Zoom
    \item Go to meeting
    \item Slack video calling
    \item Skype
    \item Facetime
    \item WhatsApp video calling
    \item Plain old telephone system
    \item Other (please specify)
    \end{inparaenum}

  \item  Do these tools help you to communicate and collaborate (e.g share
    knowledge, resolve issues quicker etc.) more efficiently while working
    from home compared to in an office environment?
    \begin{inparaenum}[$\square$]
    \item Much more efficient
    \item Somewhat more efficient
    \item About the same
    \item Somewhat less efficient.\\
    \item Much less efficient
    \end{inparaenum}

  \item  What technology do you use frequently to communicate asynchronously with the global or local team (select all that
    apply)?
    \begin{inparaenum}[$\square$]
    \item Microsoft Teams chat
    \item Zoom chat
    \item Microsoft chat
    \item SMS
    \item Slack
    \item Skype chat
    \item WhatsApp
    \item Git
    \item Confluence
    \item Jira
    \item Email
    \item Other (please specify)
    \end{inparaenum}

  \item  Do these tools help you to communicate and collaborate (e.g share
    knowledge, resolve issues quicker etc.) more efficiently while working
    from home compared to in an office environment?
    \begin{inparaenum}[$\square$]
    \item Much more efficient
    \item Somewhat more efficient
    \item About the same
    \item Somewhat less efficient
    \item A lot less efficient
    \end{inparaenum}

  \item  How has the {number} of meetings you attend changed compared to
    pre-pandemic?
    \begin{inparaenum}[$\square$]
    \item Many more meetings
    \item Somewhat more meetings
    \item About the same
    \item Somewhat fewer meetings
    \item A lot fewer meetings
    \end{inparaenum}

  \item  How has the {length} of meetings you attend changed compared to
    pre-pandemic?
    \begin{inparaenum}[$\square$]
    \item Much longer meetings
    \item Somewhat longer meetings
    \item About the same
    \item Somewhat shorter meetings
    \item Much shorter meetings
    \end{inparaenum}

  \item  Have you changed your work from home technology to meet changing needs
    in the pandemic?
    \begin{inparaenum}[$\square$]
    \item Reconfigure/make space for home office
    \item Increase broadband speed
    \item Obtain larger/multiple monitors
    \item Upgrade home computer speed, memory, or storage
    \item Better camera/microphone
    \item Other (please specify):
    \end{inparaenum}

  \end{asparaenum}

  \item{Process}\label{process}

  \begin{asparaenum}

  \item  How has working from home impacted your interaction with the
    distributed team?
    \begin{inparaenum}[$\square$]
    \item Much more interaction
    \item Somewhat more interaction
    \item About the same
    \item Somewhat less interaction
    \item Much less interaction
    \end{inparaenum}

  \item  Do any of the following software development practices help you
    collaborate with others remotely?
    \begin{inparaenum}[$\square$]
    \item Daily standup
    \item Backlog refinement
    \item Sprint retrospective
    \item Sprint review \& demo
    \item PI (program increment) planning.\\
    \item Communities of practice
    \end{inparaenum}

  \item  Looking back to the beginning of the pandemic (March-April 2020), how
    did your productivity at the beginning of the pandemic compare to
    pre-pandemic?\label{question:productivity-begin}
    \begin{inparaenum}[$\square$]
    \item Much more productive at beginning of pandemic than pre-pandemic
    \item Somewhat more productive at beginning of pandemic than pre-pandemic
    \item About the same
    \item Somewhat less productive at beginning of pandemic than pre-pandemic
    \item Much less productive at beginning of pandemic than pre-pandemic
    \item Don't know -- joined \Ocuco during the pandemic
    \end{inparaenum}

  \item  How is your productivity \emph{now} compared to pre-pandemic?
    \begin{inparaenum}[$\square$]\label{question:productivity-now}
    \item Much more productive now than pre-pandemic
    \item Somewhat more productive now than pre-pandemic
    \item About the same
    \item Somewhat less productive now than pre-pandemic
    \item Much less productive now than pre-pandemic
    \item Don't know -- joined \Ocuco during the pandemic
    \end{inparaenum}

  \item  How is your productivity \emph{now} compared to the beginning of the
    pandemic (March-April 2020)?\label{question:productivity-begin-now}
    \begin{inparaenum}[$\square$]
    \item Much more productive now than at the beginning of the pandemic
    \item Somewhat more productive now than at the beginning of the pandemic
    \item About the same now than at the beginning of the pandemic
    \item Somewhat less productive now than at the beginning of the pandemic
    \item Much less productive now than at the beginning of the pandemic
    \item Don't know -- joined \Ocuco recently
    \end{inparaenum}

  \item  How effective are \Ocuco's pre-pandemic distributed development
    processes and practices for working from home?
    \begin{inparaenum}[$\square$]
    \item Very effective
    \item Somewhat effective
    \item Neither effective nor ineffective
    \item Somewhat ineffective
    \item Very ineffective
    \item Don't know -- joined \Ocuco during the pandemic
    \end{inparaenum}

  \item  Is there any practice introduced for working from home that you would
    like to see implemented in the office environment post-pandemic
    (flexibility of working hours, better communication, helpful attitude,
    etc.)? Please elaborate:

  \end{asparaenum}

  \item{Wellness}\label{wellness}

  \begin{asparaenum}

  \item  During the pandemic, have you found it is more or less difficult to
    concentrate as compared to pre-pandemic?
    \begin{inparaenum}[$\square$]
    \item Much more difficult
    \item Somewhat more difficult
    \item About the same
    \item Somewhat less difficult
    \item Much less difficult
    \end{inparaenum}

  \item  During the pandemic, have you felt more or less anxiety as compared to
    pre-pandemic?
    \begin{inparaenum}[$\square$]
    \item Much more anxious
    \item Somewhat more anxious
    \item About the same
    \item Somewhat less anxious
    \item Much less anxious
    \end{inparaenum}

  \item  During the pandemic, how often do you engage in vigorous activities
    (like running or HIIT) compared to pre-pandemic?
    \begin{inparaenum}[$\square$]
    \item Much more often
    \item Somewhat more often
    \item About the same
    \item Somewhat less often
    \item Much less often
    \end{inparaenum}

  \item  During the pandemic, how often do you engage in moderate activities
    (other than walking) compared to pre-pandemic?
    \begin{inparaenum}[$\square$]
    \item Much more often
    \item Somewhat more often
    \item About the same
    \item Somewhat less often
    \item Much less often
    \end{inparaenum}

  \item  During the pandemic, how often do you engage in walking compared to
    pre-pandemic?
    \begin{inparaenum}[$\square$]
    \item Much more often
    \item Somewhat more often
    \item About the same
    \item Somewhat less often
    \item Much less often
    \end{inparaenum}

  \item  Overall, how do you feel your wellbeing has been impacted during the
    pandemic?
    \begin{inparaenum}[$\square$]
    \item Very positive impact
    \item Somehat positive impact
    \item No impact
    \item Somewhat negative impact
    \item Very negative impact
    \end{inparaenum}

  \item  Any other comment?

  \end{asparaenum}
\end{asparaenum}
 }

\end{document}